\documentclass[preprint,superscriptaddress,floats,showpacs,10pt]{revtex4-1}

\usepackage{graphicx,epsfig}
\usepackage{amsfonts,amsmath,amssymb,amsthm,amscd}
\usepackage[T2A]{fontenc}
\usepackage{color}
\usepackage{dcolumn}
\usepackage{bm}
\usepackage{array}
\usepackage{float}
\begin{document}
\title{Helium-3 and Helium-4 acceleration by high power laser pulses for hadron therapy}

\author{S. S. Bulanov}
\affiliation{Lawrence Berkeley National Laboratory, Berkeley, California 94720, USA}

\author{E. Esarey}
\affiliation{Lawrence Berkeley National Laboratory, Berkeley, California 94720, USA}

\author{C. B. Schroeder}
\affiliation{Lawrence Berkeley National Laboratory, Berkeley, California 94720, USA}

\author{W. P. Leemans}
\affiliation{Lawrence Berkeley National Laboratory, Berkeley, California 94720, USA}
\affiliation{Department of Physics, University of California, Berkeley, California 94720, USA}

\author{S. V. Bulanov}
\affiliation{Kansai Photon Science Institute, JAEA, Kizugawa, Kyoto 619-0215, Japan}
\affiliation{Prokhorov Institute of General Physics, Russian Academy of Sciences, Moscow
119991, Russia}
\affiliation{Moscow Institute of Physics and Technology, Dolgoprudny, Moscow region, 141700 Russia}

\author{D. Margarone}
\affiliation{Institute of Physics ASCR, v.v.i. (FZU), ELI-Beamlines Project, 182 21 Prague, Czech Republic}

\author{G. Korn}
\affiliation{Institute of Physics ASCR, v.v.i. (FZU), ELI-Beamlines Project, 182 21 Prague, Czech Republic}

\author{T. Haberer}
\affiliation{Heidelberger Ionenstrahl-Therapie Centrum (HIT), D -69120 Heidelberg, Germany}

\date{\today}

\begin{abstract}
The laser driven acceleration of ions is considered a promising candidate for an ion source for hadron therapy of oncological diseases. Though proton and carbon ion sources are conventionally used for therapy, other light ions  can also be utilized. Whereas carbon ions require 400 MeV per nucleon to reach the same penetration depth as 250 MeV protons, helium ions require only 250 MeV per nucleon, which is the lowest energy per nucleon among the light ions. This fact along with the larger biological damage to cancer cells achieved by helium ions, than that by protons, makes this species an interesting candidate for the laser driven ion source.   Two mechanisms (Magnetic Vortex Acceleration and hole-boring Radiation Pressure Acceleration) of PW-class laser driven ion acceleration from liquid and gaseous helium targets are studied with the goal of producing 250 MeV per nucleon helium ion beams that meet the hadron therapy requirements.  We show that He$^3$ ions, having almost the same penetration depth as He$^4$ with the same energy per nucleon, require less laser power to be accelerated to the required energy for the hadron therapy.                    
\end{abstract}

\maketitle

\section{Introduction.}

{\noindent} Beams of high energy particles, initially produced for the study of fundamental processes of particle physics, found applications in different areas of science and technology. In particular, the medical applications of such beams gave rise to the field of radiation therapy \cite{HT review}. Hadron therapy, the part of radiation therapy that uses ion beams has a number of advantages, such as a  dose profile with a high dose deposition near the beam stopping point (Bragg peak). It provides a conformal dose deposition treatment, in which the radiation dose is matched to the tumor, while the radiation induced damage to the normal tissue is limited. It is known that the light ions (such as protons) allow for a more precise dose delivery in the longitudinal direction, but experience more side scattering than heavier ions (such as carbon). Heavier ions, additionally, generate higher radiation induced damage (RBE - Relative Biological Effectiveness). This damage is hard to compensate by the cell repair mechanisms \cite{lightionreview}, which is critically important for the treatment of radioresistant tumors. Heavier ions, however, generate a tail in the energy transfer distribution near the beam stopping point mainly due to the fragmentation of ions in the course of interaction with the atom nuclei in the tumor. The fragmentation may lead to the unwanted irradiation of surrounding healthy tissue. This effect can be mitigated by using the Intensity-Modulated Radiation Therapy. Mostly protons and carbon ions are used at hadron therapy facilities, however different species of ions have also been tested. One can consider optimizing ion species to maximize the RBE for the necessary accuracy of the dose delivery.

Moreover, hadron therapy may be optimized not only by using different ion species but also by using different ion sources. Presently all hadron therapy facilities in operation are based on the conventional accelerator technology making use of accelerators and complex ion beam transport systems (gantry) for beam delivery in the treatment room \cite{HT review}. It is widely discussed whether a different source of ion beams can be utilized to reduce the cost and size of these facilities. One of the possibilities is the use of high power laser systems \cite{hadron therapy}, which, in principle, exhibit lower construction and operation costs, and moreover would require smaller beam transport system, leading to a more compact facilities. Due to the rapid evolution of laser technology it became possible to use such systems for efficient acceleration of charged particles \cite{electron acceleration review, ELI, PW, ion acceleration review}. The theoretical and computer modeling results show that PW-class lasers, which are already in operation \cite{PW} and are being constructed \cite{ELI}, are able to produce ions with the energy high enough to meet the requirements for hadron therapy \cite{ion acceleration review}. Experimentally proton beams with the maximum energy of 45 MeV for short (tens of femtoseconds) \cite{45 MeV TNSA, 45 MeV RPA} and 160 MeV for long (hundreds of femtoseconds) \cite{160 MeV BOA} laser pulses were observed. The proof-of-principle experiments testing the ability of the laser produced proton beams to damage cancer cells were recently carried out \cite{hadron therapy experiment}. However the generation of a laser based ion source that would meet the maximum energy, energy spread, number of particles per bunch, and repetition rate requirements \cite{hadron therapy review, newsletter56} simultaneously is still a challenge both theoretically and experimentally.                 

In this paper we study  different mechanisms of generating a laser driven source 
of helium ions. Usually either protons or carbon ions are considered for medical applications 
of laser acceleration \cite{hadron therapy review, newsletter56}. 
Here we consider an intermediate ion species that may be more beneficial 
for therapy due to higher biological damage than that of protons and lower laser energy input 
than that for carbon ions. There are some indications that helium ions 
demonstrate higher biological damage and good dose localization with the lower dose deposition 
at the beam entry \cite{He med}. The penetration depth of an ion beam is determined by its initial energy. For a $\delta$-shaped initial ion energy distribution, $f(0,\mathcal{E})=\delta(\mathcal{E}-\mathcal{E}_{a})$, the location of the Linear Energy Transfer (LET) maximum is $x_{max}=\mathcal{E}_{a}^{2}/\eta$ \cite{hadron therapy review}, where $\eta=4\pi e^4m_a Z_u Z_a^2/m_e$, here $Z_a, m_a$ are the charge state and mass 
of the ion; $Z_u$ is the charge state of the target; $m_e$ and $e$ are electron mass 
and charge; $\mathcal{E}_{a}$ is the initial ion energy. 
This relation allows us to estimate how much energy per nucleon is needed 
for an ion beam to have the same penetration depth as a proton beam with initial energy of $\mathcal{E}_p$:
\begin{equation}
\left(\mathcal{E}_{a}/A_a\right)=\mathcal{E}_{p}\left(Z_a/A_a^{1/2}\right),
\end{equation}
here $A_a$ is the ion atomic number. For all stable isotopes of elements from helium to carbon the factor $Z_a/A_a^{1/2}$ ranges from 1 to 1.73, being the smallest for $He^4$ and the largest for carbon.  
The isotope $He^3$ has $Z_{He^3}/A_{He^3}^{1/2}=1.15$. This makes $He^3$ and $He^4$ ions interesting candidates 
for hadron therapy, since they have the lowest energy per nucleon requirement among 
the ions heavier than protons, and deliver higher biological damage. 

In order to be practical the laser driven ion source should be able to deliver helium ions with the energy up to 250 MeV/u with a short pulse PW-class laser system and the target should be replenishable to allow for the high repetition rate. In the case of Helium it is reasonable to consider two types of targets: (i) liquid and (ii) gaseous. The first target can be realized as a liquid jet or a wall of a liquid helium bubble. The gas target may be realized as either a very narrow high density gas jet or a shock launched inside a low density gas jet. In principle, the available laser power and the properties of the target make it possible to choose the regime of interaction that would allow to optimize the parameters of the produced ion beam. There are several basic laser ion acceleration mechanisms: (i) Target Normal Sheath Acceleration (TNSA) \cite{TNSA}, (ii) Coulomb Explosion (CE) \cite{CE}, (iii) Radiation Pressure Acceleration (RPA) \cite{RPA}, (iv) Magnetic Vortex Acceleration (MVA) \cite{MVA_first,MVA}, and (v) the Shock Wave Acceleration (SWA) \cite{SWA}. There are also several mechanisms that through either modification or combination of some of the basic regimes enhance the maximum ion energy, number of accelerated ions, or improve their spectrum. For example, the Burn-out-Afterburner (BOA) \cite{BOA}, which employes an enhanced TNSA, and Directed Coulomb Explosion (DCE) \cite{DCE}, which is the combination of RPA and CE, are such composite mechanisms. Also the use of composite targets (consisting of low density and high density parts) was proposed in a number of papers to either inject the ions into accelerating fields or to enhance the interaction of the laser pulse with the high density part of the target \cite{TeV_protons, Macchi, max energy}. Most of the above mentioned mechanisms were shown theoretically \cite{ion acceleration review} to produce high energy ion beams from either ultra-thin (tens or hundreds of nanometers) solid density targets or employing long pulse lasers, which can not be achieved in the case of liquid and gaseous helium targets and short laser pulses. Thus we consider two mechanisms of laser ion acceleration relevant to these targets and able to provide necessary  peak energy. In the case of a very thin liquid target it is Radiation Pressure Acceleration (RPA) \cite{RPA,RPA_next,RPA exp}, and in the case of a gas target it is Magnetic Vortex Acceleration (MVA) \cite{MVA_first,MVA,MVA exp}. 

\section{Liquid Helium target. Radiation PressureAcceleration.}

{\noindent} The interaction of a laser pulse with a liquid helium target can be viewed 
as "hole-boring" \cite{hole-boring RPA}, which is RPA of a density modulation inside an extended target. 
In principle, the energy gain in this regime can be estimated by assuming 
that there exists a stationary solution ($d\beta/dt=0$, where $\beta$ is the velocity 
of the laser plasma interface). The position of the laser plasma interface 
is then determined by the balance between the ion momentum flux and the flux of the 
EM wave momentum. This gives \cite{slow wave} for the velocity of this interface:
\begin{equation}\label{HB}
\beta=\frac{\sqrt{4\beta_g B_E^2+(1-\beta_g^2)B_E^4}-B_E^2(1+\beta_g^2)}{2(1-\beta_g B_E^2)},
\end{equation}
where $B_E=\left(E^2/2\pi n_e m_{He}c^2\right)^{1/2}$, $\beta_g$ 
is the group velocity of the laser pulse, and $n_e$ is the electron density. 
Here we took into account the fact that a tightly focused laser pulse has a group 
velocity smaller than the vacuum light speed, which ultimately 
limits the maximum achievable ion energy \cite{max energy,slow wave}. 
For a 1 PW laser pulse focused down to 1.5$\mu$m focal spot at helium 
plasma with the density of 28$n_{cr}$, Eq. (\ref{HB}), where the laser 
pulse EM intensity is averaged over time and space, gives an energy gain 
of ${\cal E}_{He}\simeq 0.6$ GeV, which is $\sim 150$ MeV per nucleon. 
Here $n_{cr}=\omega^2 m_e/4\pi e^2$ is the critical plasma density and 
$\omega$ is the laser frequency. The scaling for the energy 
gain with the increase of the laser power from Eq. (\ref{HB}) is ${\cal E}_{He}\sim P^{1/2}$.
  
\begin{figure}[tbp]
\epsfxsize8cm\epsffile{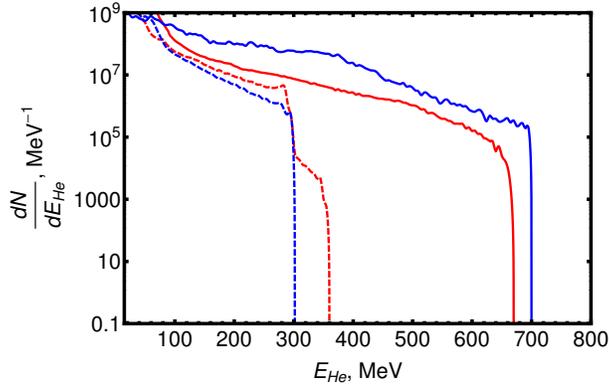}
\caption{The spectra of helium-4 (blue curves) and helium-3 (red curves) ions accelerated from a 6$\lambda$ 
liquid target by 1 PW (dashed curves) and 2 PW (solid curves) laser pulses in the forward direction, i.e., $p_{He}>0$.}
\label{FIG2}
\end{figure}

In what follows we present the results of 2D PIC simulations of a PW-class laser 
interaction with a He target with thickness $\sim 6\lambda$, where $\lambda$ is the laser wavelength, and electron density of $28 n_{cr}$ for He$^4$ and $13 n_{cr}$ for He$^3$, which corresponds to the liquid Helium density near the boiling point (4.7 K). The simulations are performed using code REMP \cite{REMP}. 
The simulations are set up as follows: the simulation box is 
20$\lambda$ x 10$\lambda$, dt=0.0025$\lambda$, and dx=dy=0.005$\lambda$, 
chosen in order to resolve the skin depth. The laser is introduced at the 
left boundary with Super Gaussian longitudinal and transverse profiles, $\exp\left\{-(y/w)^8-[2(x-t)/\tau]^8\right\}$, duration of $\tau=30$ fs, and spot size of $w=2\lambda$. The utilization of such laser profile was shown \cite{DCE} to be beneficial for enhancing the maximum ion energy. The laser pulse is focused 8$\lambda$ from the left border with f/D=2, giving the laser spot size at focus of about 1.5$\lambda$. The laser pulse bores a hole in the liquid target, accelerating helium ions at the laser-plasma interface, which is a characteristic feature of the hole-boring RPA mechanism. The spectra of He$^3$ and He$^4$ are exponentially decaying with a cutoff energy of 300 MeV for 1 PW laser pulse and $\sim 700$ MeV for 2 PW laser pulse (see Fig. \ref{FIG2}). Only in the 2 PW case the helium ion energy partly overlaps with the required for hadron therapy energy range of (240 MeV,860 MeV) for He$^3$ and (280 MeV, 1 GeV) for He$^4$. However the fast decay of the spectra leads to a very small number of particles near the cutoff energy: $3\times 10^7$ in the energy interval $\{0.98 \mathcal{E}_{He}, 1.02 \mathcal{E}_{He}\}$, with $\mathcal{E}_{He}=600$ MeV. The utilization of thin liquid targets will require 
a high level of control over the laser prepulse to prevent the significant modification 
of the target before the arrival of the main pulse. In some cases the existence of 
the prepulse can boost the acceleration mechanism by providing more 
favorable conditions for the laser pulse coupling to the target, 
however high level of control over the prepulse is mandatory in this case \cite{Prepulse Esirkepov}. 

\section{Gaseous target. Magnetic Vortex Acceleration.}

{\noindent}In the MVA regime, during the laser pulse propagation in near critical density (NCD) plasma the ponderomotive force of the laser expels electrons and ions in the transverse direction. The time of ion response is much longer than that of electrons, which leads to a formation of a positively charged region just after the front of the laser pulse. Thus the electrons move under the action of the laser pulse EM field and the field of a remaining ion core. 
This results in the formation of a channel in electron density with high density walls. After the laser pulse passes, the ions also expand in the transverse direction forming a channel in ion density. 
The propagation of an intense laser pulse in the relativistically 
underdense plasma can be approximated by the propagation of an EM wave in a waveguide \cite{MVA}. 
The radius of the self-generated channel can be estimated from balancing the electron energy gain 
from the laser pulse EM field and from field generated by the positive charge of ions in the channel: 
$R_{ch}=\left(a_{ch} n_{cr}/n_e\right)^{1/2}\lambda/\pi$. The relation between the amplitude of laser pulse vector-potential in the channel, $a_{ch}$, and the laser pulse power, $P$, is
\begin{equation}\label{a_ch}
a_{ch}=\left(\frac{2}{K}\frac{P}{P_c}\frac{n_e}{n_{cr}}\right)^{1/3},
\end{equation}
where $P_c=2 m_e^2 c^5/e^2=17$ GW is a characteristic power for relativistic self focusing \cite{SUN}. The factor $K=0.074$  comes from the integration over transverse coordinates and the duration of the laser pulse. The radius of the channel in terms of the laser pulse power is
\begin{equation} \label{channel radius}
R_{ch}=\frac{\lambda}{\pi}\left(\frac{n_{cr}}{n_e}\right)^{1/3}\left(\frac{2}{K}\frac{P}{P_c}\right)^{1/6}.
\end{equation}

In the MVA regime the acceleration is achieved by an electric field at the back of the target, which is generated by, and is of order of, the magnetic field ($B_{ch}$) of electrons, accelerated by the laser pulse in plasma in the forward direction, as they exit the target. This EM field associated with the electron current leaving the target expands along the back surface of the target, however the magnetic field flux is conserved. 
The maximum energy gain of a charged particle in such an expanding field is
$\mathcal{E}_{He}\sim e Z_{He} B_{ch}R_{ch}$.  

Let us assume that the electron current consists of all the electrons that initially 
were in the volume $\pi R_{ch}^3$, which is  the volume of a cavity left 
in the wake of a laser pulse, ensuring that the fields of electrons and ions are compensated 
outside the cavity and the strong fields exist only inside it. 
As the electrons move through the charged ion background they experience a plasma 
lensing effect \cite{plasma lens}, i.e., they are pinched towards the central axis 
and the radius of the electron beams is determined from the balance of the transverse 
electric field of the ion column, $E_i=2\pi e n_e R_b$, which in the reference 
frame of an e-beam is $E_i^\prime=\gamma_e E_i$, and the self field of an electron beam, 
which is equal to $E_e^\prime=2\pi e n_e^\prime R_b$. From the condition 
$E_i^\prime=E_e^\prime$ we obtain $R_b=R_{ch}/\gamma_e$. Here $\gamma_e$ 
is the Lorentz factor of the bulk of electrons accelerated forward. 
In the regime of a laser pulse interaction with a NCD plasma the 
electrons are continuously injected in the cavity behind the laser pulse 
leading to a formation of an electron current, which is dominated by low energy electrons.

The characteristic energy of these electrons is the injection energy, 
which can be obtained from the condition on the electron velocity to be equal to the 
group velocity of an EM pulse propagating in a waveguide of radius $R_{ch}$: $
\gamma_e=(\sqrt{2}/1.84)\left(2 P/ K P_c\right)^{1/6}\left(n_{cr}/n_e\right)^{1/3}$.        
The magnetic field, generated by these electrons, is $B_{ch}(R_b)=2\pi e n_e R_{ch}\gamma_e^2$ at $r=R_b$. Then the energy gain of an ion in the electric field at the back of the target is
\begin{equation}\label{MVA energy gain}
\mathcal{E}_{He}=m_ec^2\,2\pi^2 Z_{He}\left(n_e/n_{cr}\right)\left(R_{ch}/\lambda\right)^4,
\end{equation}
which gives an upper estimate for the ion energy gain in the MVA regime. For example, for a 1 PW laser pulse and $n_e=3 n_{cr}$ $\mathcal{E}_{He}\approx 3$ GeV. From Eq. (\ref{MVA energy gain}) we see that the ion energy scales with the laser pulse power as $P^{2/3}$. The scaling was derived assuming the optimal match between the laser pulse and the plasma, i.e., the target length is equal to the laser pulse depletion length. We should mention here that MVA regime is relevant for ion acceleration from targets with densities being not far from  critical \cite{MVA}. For lower densities, laser pulse filamentation and Langmuir wave generation prevent efficient channel generation. At higher densities other mechanisms of laser ion acceleration come into play and the target thickness, which is equal to the depletion length, is of the order the pulse length or smaller, which is outside the regime of MVA  applicability.

\begin{table*}
\begin{tabular}[c]{||c||c||c||c||}
\hline   & $n_e=n_{cr}$ & $n_e=2n_{cr}$ & $n_e=3n_{cr}$  \\ 
& \begin{tabular}[b]{c|c} \hline He$^3$ & He$^4$ \end{tabular} & \begin{tabular}[b]{c|c} \hline He$^3$ & He$^4$ \end{tabular}& \begin{tabular}[b]{c|c} \hline He$^3$ & He$^4$ \end{tabular} \\ \hline \hline
  P=500 TW & \begin{tabular}[b]{c|c} $8\times 10^7$ & 0 \\ 380 MeV & 190 MeV  \end{tabular} & \ \begin{tabular}[b]{c|c} $3\times 10^8$ & 0 \\ 570 MeV & 230 MeV  \end{tabular}  & \begin{tabular}[b]{c|c} $1\times 10^8$ & $3\times 10^7$ \\ 400 MeV & 290 MeV  \end{tabular} \\ \hline
  P=1 PW & \begin{tabular}[b]{c|c} $6\times 10^8$ & $2\times 10^8$ \\ 520 MeV & 380 MeV  \end{tabular} & \begin{tabular}[b]{c|c} $9\times 10^8$ & $4\times 10^8$ \\ 670 MeV & 450 MeV  \end{tabular} & \begin{tabular}[b]{c|c} $8\times 10^8$ & $3\times 10^8$ \\ 830 MeV & 530 MeV  \end{tabular}  \\ \hline
  P=2 PW & \begin{tabular}[b]{c|c} $1.6\times 10^9$ & $1.5\times 10^9$ \\ 770 MeV & 550 MeV  \end{tabular} & \begin{tabular}[b]{c|c} $2.6\times 10^9$ & $1.2\times 10^9$ \\ 930 MeV & 650 MeV  \end{tabular} & \begin{tabular}[b]{c|c} $1.5\times 10^9$ & $8\times 10^8$ \\ 1100 MeV & 880 MeV  \end{tabular}  \\ \hline
\end{tabular}
\caption{\label{<Table>} 
The number of He$^3$ and He$^4$ ions in the energy interval 
$\{0.98 {\cal E}_{max},1.02{\cal E}_{max}\}$, where ${\cal E}_{max}=250 A_{He}$ MeV, and the maximum helium ion energy per nucleon vs laser power and density of the target.}
\end{table*} 
\begin{figure}[tbp]
\epsfxsize8cm\epsffile{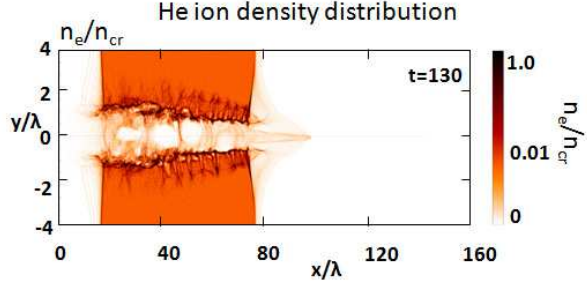}
\caption{The distribution of He$^4$ ion density in the course 
of a 1 PW laser pulse interaction with a 60$\lambda$ NCD He$^4$ target at $t=130\times 2\pi/\omega$.}
\label{FIG3}
\end{figure} 
\begin{figure}[tbp]
\epsfxsize7cm\epsffile{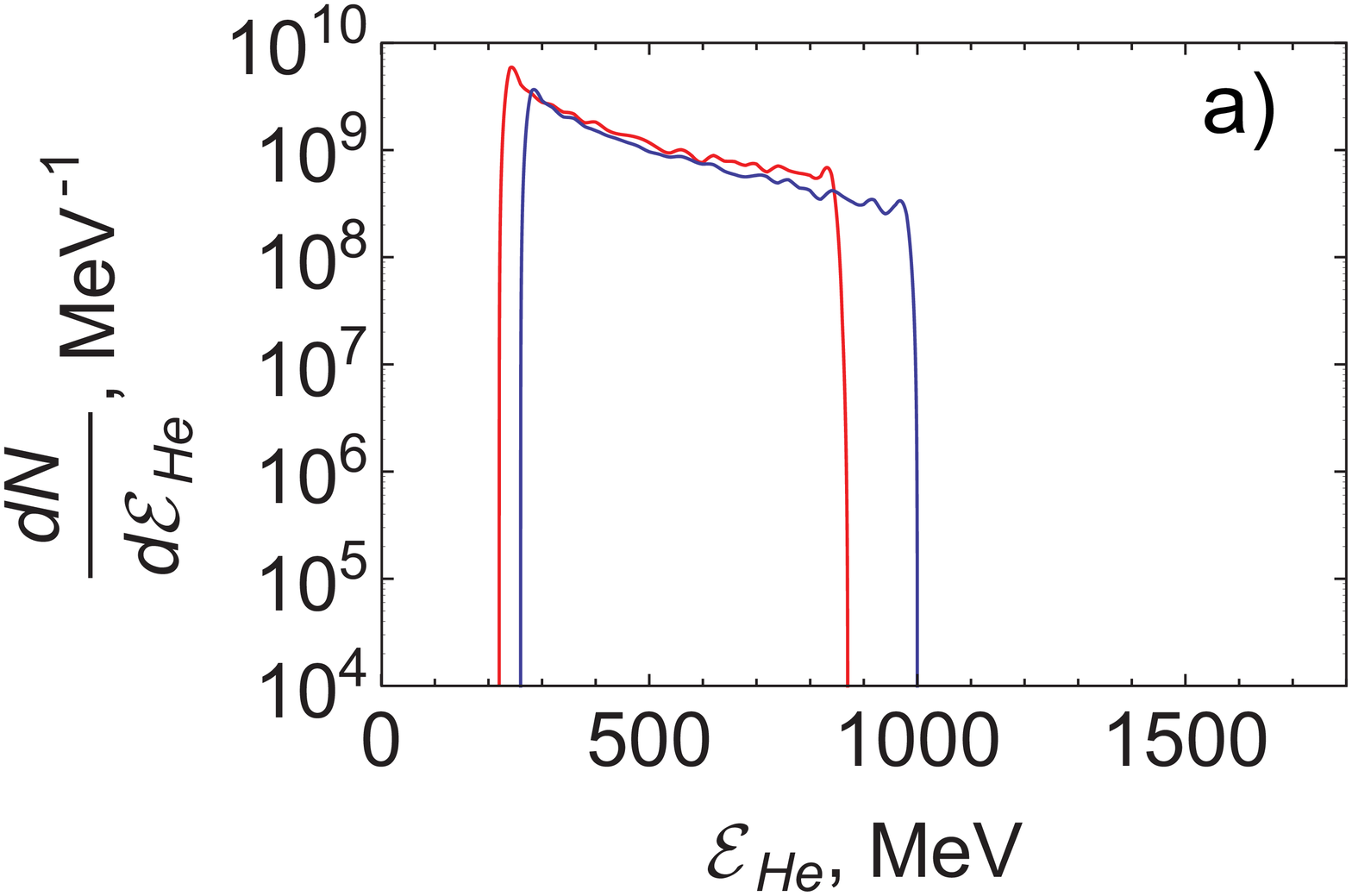}
\epsfxsize7cm\epsffile{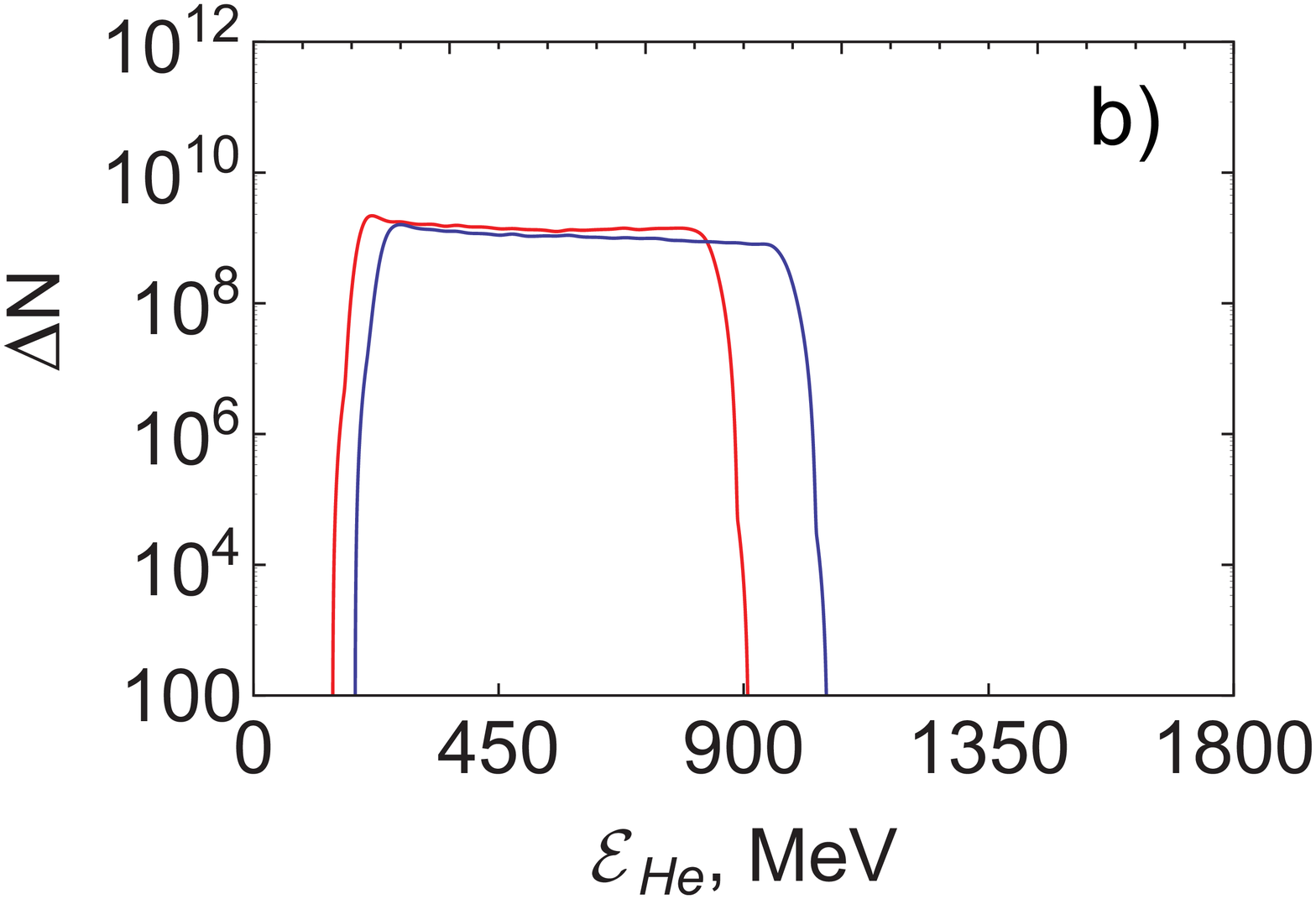}
\caption{a) The spectrum of helium-3 (red) 
from 240 MeV to 860 MeV and helium-4 (blue) 
ions from 280 MeV to 1 GeV accelerated in the forward direction; 
b) The number of helium-3 (red) and helium-4 (blue) ions per energy interval, 
$0.98 \mathcal{E}_{He}<\mathcal{E}_{He}<1.02\mathcal{E}_{He}$. 
The laser pulse power is 1 PW, the target thickness is 60 $\lambda$.}
\label{FIG4}
\end{figure}     

In what follows we present the results of 2D PIC simulations \cite{REMP} 
of a PW-class laser interaction with He$^4$ and He$^3$ targets with thickness (30 - 100)$\lambda$ 
and electron density of (1-3)$n_{cr}$, which corresponds to the density of a high pressure Helium gas jet. 
The simulations are set up as follows: the simulation box is 150$\lambda$ x 100$\lambda$, dt=0.05$\lambda$, 
and dx=dy=0.1$\lambda$. The laser is introduced at the left boundary with Super Gaussian longitudinal and transverse profiles, $\exp\left\{-(y/w)^8-[2(x-t)/\tau]^8\right\}$, duration of 30 fs, and spot size of $w=4\lambda$. The laser pulse is focused 16$\lambda$ from the left border with f/D=2, giving the laser spot size 
at focus of about 1.5$\lambda$. The distribution of helium ions during the interaction is shown in Fig. \ref{FIG3}. 
A thin ion filament formed along the laser propagation axis is clearly seen, 
as the ions from it are accelerated by the charge separation electric field formed at the back of the target. 
The spectra of He$^3$ and He$^4$ ions are shown in Fig. \ref{FIG4}a for the interval of energies 
relevant to hadron therapy: 280 MeV$<\mathcal{E}_{He^4}<$1 GeV and 240 MeV$<\mathcal{E}_{He^3}<$ 860 MeV. Such energy selection can be achieved by a system based on permanent dipoles as shown in Ref. \cite{energy selection}.   
The slow decay of the spectrum in the interval relevant to therapy needs makes 
it possible to utilize different energy beamlets corresponding to different penetration depths simultaneously, 
thus greatly increasing the effectiveness of the laser driven ion source. 
The number of ions per energy interval $\{0.98\mathcal{E}_{He},1.02 \mathcal{E}_{He}\}$ is shown in Fig. \ref{FIG4}b. We see that it demonstrates a very slow decay with a difference between 
the maximum and minimum number less than two times.

We summarize the results of PIC simulations in Table I. The maximum energy per nucleon varies from 400 MeV to 1.1 GeV per nucleon for He$^3$ and from 190 MeV to 880 MeV per nucleon for He$^4$. 
He$^3$ ions have higher maximum energy per nucleon for all considered density and 
laser power values, yielding a higher number of particles in the energy interval, 
$\{0.98\mathcal{E}_{max},1.02\mathcal{E}_{max}\}$. According to simulation results, 500 TW laser power can be considered a threshold of therapeutic beam production with a number of particles per bunch of about $10^8$ 
in the case of He$^3$. Further increase of the laser power up to 2 PW shows 
an increased production of He ions with a number of particles per bunch reaching  
$2\times 10^9$. 

\section{Conclusions.}

{\noindent}We considered a laser driven helium ion source for the hadron therapy. 
Two types of targets were studied: a liquid target (accelerated using hole-boring RPA) and a gaseous target (accelerated using RPA). We found that the liquid helium target requires much higher laser power to produce 
ion beams that may be of interest for therapy than the gaseous target, namely, 2 PW vs 500 TW, and produces significantly less particles per energy bin, namely, $10^7$ vs $10^9$. 
Moreover the spectrum of the helium ions produced through the MVA regime allows 
for the multiple energy bin extractions to simultaneously treat the parts 
of tumor corresponding to different penetration depths. 
The number of particles per energy bin is about $10^9$, 
which combined with the 1 Hz laser repetition rate meets 
the hadron therapy requirements regarding the total dose and 
the procedure duration. We also investigated the feasibility of using He$^3$ isotope 
for the laser driven ion source. He$^3$, having almost 
the same penetration depth as He$^4$ with the same energy per nucleon, requires 
less laser power to be accelerated to the required energy for hadron therapy. This He$^3$ advantage should be weighted against the relative biological damage of He$^3$ and He$^4$ ions, but this optimization lies outside the scope of this paper.               

\acknowledgements
We thank T. Zh. Esirkepov for providing REMP code for simulations. 
We acknowledge support from the Office of Science of the US DOE under 
Contract No. DE-AC02-05CH11231 and No. DE-FG02-12ER41798 and the Ministry of Education, Youth and Sports of the Czech Republic (ELI-Beamlines reg. No. CZ.1.05/1.1.00/02.0061)

\end{document}